\documentclass[10pt,letterpaper]{article}
\usepackage{opex3}
\usepackage{subfigure}
\usepackage{cite}
\usepackage{graphics}

\begin{document}




\title{Ramsey spectroscopy of high-contrast CPT resonances with push-pull optical pumping in Cs vapor}

\author{X. Liu,$^1$ J-M. M\'erolla,$^1$ S. Gu\'erandel,$^2$  E. de Clercq$^2$ and R. Boudot$^1$$^{\ast}$}

\address{$^1$FEMTO-ST, CNRS, 26 chemin de l'Epitaphe 25030 Besancon Cedex, France.\\
$^2$LNE-SYRTE, Observatoire de Paris, CNRS, UPMC, 61 avenue de l'Observatoire, 75014 Paris, France.\\
}




\begin{abstract} We report on the detection of high-contrast and narrow Coherent Population Trapping (CPT) Ramsey fringes in a Cs vapor cell using a simple-architecture laser system. The latter allows the combination of push-pull optical pumping (PPOP) and a temporal Ramsey-like pulsed interrogation. An originality of the optics package is the use of a single Mach-Zehnder electro-optic modulator (MZ EOM) both for optical sidebands generation and light switch for pulsed interaction. Typical Ramsey fringes with a linewidth of 166 Hz and a contrast of 33 \% are detected in a cm-scale buffer-gas filled Cs vapor cell. This technique could be interesting for the development of high-performance and low power consumption compact vapor cell clocks based on CPT.
\end{abstract}



\section{Introduction}
The development of CPT atomic clocks \cite{Vanier:APB:2005,Knappe:Elsevier:2010} of high frequency stability requires to detect a narrow atomic resonance with the highest signal to noise ratio. In traditional CPT clocks, atoms interact with circularly polarized laser light. Numerous atoms are then trapped in extreme Zeeman sublevels where they do not contribute to the magnetic field insensitive 0-0 clock transition. This technique limits to a few percent the contrast, defined as the signal height to background ratio, of the detected resonance. To circumvent this issue, different optimized double-lambda CPT pumping schemes such as push-pull optical pumping (PPOP) \cite{Jau:PRL:2004}, the use of polarized counterpropagating waves of $\sigma^+ - \sigma^-$ field configuration in small cells \cite{Taiche:2004}, \emph{lin-per-lin} \cite{Zanon:PRL:2005} or \emph{lin-par-lin} \cite{Taichenachev:JETP:2005} were proposed to maximize the number of atoms in the clock levels. More recently, Yudin et al. proposed a non-standard spectroscopic technique using a feedback control of the input probe field parameters to detect atomic resonances with high contrast \cite{Yudin:2013}.\\
\indent The detection of high contrast resonances with optimized CPT interaction schemes is known to be improved with high operating laser intensities \cite{Jau:PRL:2004, Zanon:PRL:2005, Zibrov:PRA:2010}. Nevertheless, in continuous regime, high intensities broaden the CPT resonance line, limiting the clock short-term frequency stability and increasing the light shift. To get rid of these limitations, a well-known technique is the Ramsey interference scheme using two separated oscillating fields \cite{Ramsey:PR:1959}. The first Ramsey fringes on CPT resonance were observed by Thomas \emph{et al.} \cite{Thomas}. Godone \emph{et al.} used this method to develop a pulsed rubidium CPT maser exhibiting a frequency stability of 1.2 $\times$ 10$^{-12}$ $\tau^{-1/2}$ for measurement times up to $\tau \simeq$ 10$^5$ s \cite{Godone:PRA:2006}. Zanon \emph{et al.} proposed to combine the \emph{lin per lin} interaction scheme with a temporal Ramsey-like pulsed interrogation \cite{Zanon:PRL:2005}. The detection of high-contrast CPT Ramsey fringes with greatly reduced sensitivity to light shift was demonstrated \cite{Castagna:UFFC:2009} leading to a clock short-term frequency stability of 7 $\times$ 10$^{-13}$ at 1 s \cite{Boudot:IM:2009}. Esnault \emph{et al.} reported CPT Ramsey fringes using a \textit{lin-par-lin} scheme on cold rubidium atoms. A frequency stability of 4 $\times$ 10$^{-11}$ $\tau^{-1/2}$ was obtained \cite{Esnault:2012}.  Nevertheless, these clocks remain complex and voluminous systems.\\
\indent Recently, interesting studies were proposed to implement advanced CPT clock architectures with original and simple setups \cite{Yim:RSI:2008, Yun:RSI:2011, Zhang:OE:2012, Yun:EPL:2012}. Up to date, to our knowledge, no frequency stability performances were measured with these setups. However, we consider that such investigations are interesting to make compact CPT atomic clocks suitable for industrial transfer that requires low power consumption and a reduced number of components.\\
\indent In a recent study, we investigated CPT resonances in Cs vapor cells with PPOP \cite{Liu:PRA:2013}. Resonance contrasts as high as 78 \% were demonstrated in continuous regime in cm-scale cells. However, such contrasts were obtained for high laser intensities that induce CPT line broadening. We propose here an original simple-architecture laser system combining for the first time PPOP and Ramsey spectroscopy for the detection of high-contrast and narrow Ramsey fringes. While light switching for the Ramsey pulsed interaction is currently realized in vapor cell clocks by an additional high-power consuming and polarization-sensitive acousto-optical modulator (AOM), our system uses a single MZ EOM both to generate optical sidebands and to switch on-off the light. This strategy allows to reduce the number of components and is compatible with the development of a compact and high-performance atomic clock.

\section{Experimental set-up}\label{sec:setup}
\subsection{Description}
\begin{figure}[h!]
\centering
\includegraphics[width=0.85\linewidth]{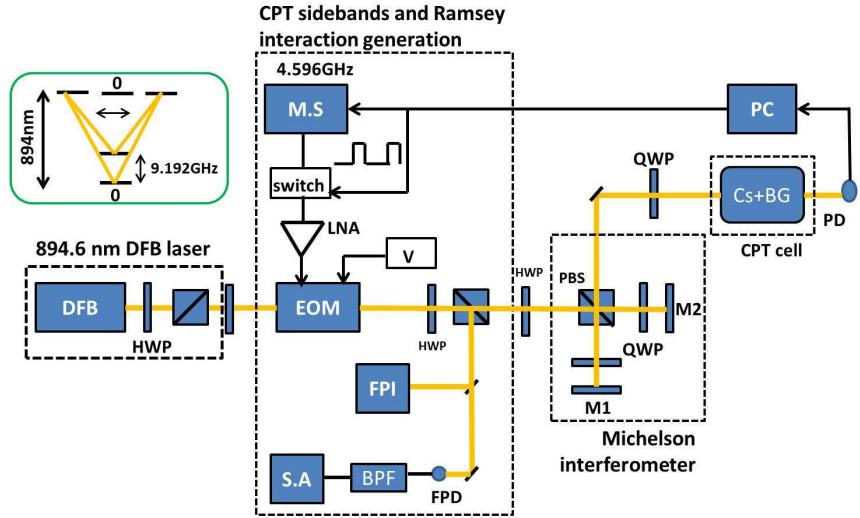}
\caption{Experimental setup. DFB: Distributed-Feedback diode laser, HWP: half-wave plate, EOM: electro-optic modulator, MS: microwave synthesizer, FPI: Fabry-Perot interferometer, BPF: bandpass filter,  S.A: microwave spectrum analyzer, PD: photodiode, FPD: fast photodiode, Cs+BG: Cs + buffer gas, PBS: polarizing beam splitter, QWP: quarter-wave plate,  M1 and M2: mirrors, LNA: low-noise amplifier. The inset shows the energy-level diagram involved in Cs vapor with PPOP. The clock transition is the ground-state 0-0 transition.}
\label{fig:figure1}
\end{figure}

Figure \ref{fig:figure1} shows the experimental set-up.

The laser source is a distributed-feedback (DFB) diode laser tuned on the Cs D$_1$ line at 894.6 nm \cite{Liu:IM:2012} and shifted from Cs atom resonance by 4.596 GHz. The laser light is injected into a polarization maintaining pigtailed MZ EOM (Photline NIR-MX800-LN-10). The latter allows both the generation of optical sidebands and the on-off light switch for the Ramsey regime. The EOM is driven at 4.596 GHz with a stabilized RF power of 22 dBm by a low noise microwave oscillator to generate first-order optical sidebands frequency-separated by 9.192 GHz. The EOM bias electrode dc voltage $V$ is actively controlled using a microwave synchronous detector technique \cite{Liu:PRA:2013} to the so-called \emph{dark point} to stabilize the carrier suppression. This carrier suppression is inherently limited by the EOM extinction ratio parameter measured to be 24 dB with our device. Both first-order sidebands exhibit equal power. The power of second-order harmonics is in correct agreement with theoretical calculations \cite{eom-ol}, about 30 dB lower than the power of first-order sidebands at the dark point. The EOM is temperature-controlled at 48$^{\circ}$C at the mK level for improved laser power stabilization at the output of the EOM \cite{Liu:2012}. At the output of the EOM, the laser is split into two directions. In the first direction, the laser light can be analyzed using a Fabry-Perot interferometer (FPI) or a fast photodiode and a spectrum analyzer. In the second direction, the modulated light of fixed linear polarization is sent through a Michelson interferometer ensemble in order to time-delay and to rotate the polarization of one half of the beam. A differential optical path $\Delta L = \lambda_0 / 4$ of about 8.1 mm, with $\lambda_0$ the clock transition wavelength, is adjusted between both arms of the interferometer. Both beams are then respectively right and left circularly polarized by a quarter-wave plate to produce PPOP  \cite{Jau:PRL:2004}. The collimated laser beam diameter is expanded to 2 cm with a combination of two convergent lenses. Laser light is then sent through a 5 cm-long and 2 cm-diameter pyrex vapor cell filled with Cs atoms and a mixture of buffer gases, Ar and N$_2$ in the pressure ratio $P_{Ar}/P_{N_2}$ = 0.4, with a total pressure $P$ of 15 Torr. The cell is temperature stabilized at the mK level at 38$^{\circ}$C and located in a static magnetic field parallel to the laser beam to raise the Zeeman degeneracy. The cell is isolated from external electromagnetic perturbations with a double layer mu-metal shield. The transmitted optical power through the cell is detected by a low noise photodiode. The resulting signal is analyzed by a computer that drives a microwave switch and the synthesizer output frequency. CPT spectroscopy is realized by scanning the local oscillator frequency around 4.596 GHz.

\subsection{Pulsed interaction with a MZ EOM}
\begin{figure}[htb]
\centering
\includegraphics[width=0.65\linewidth]{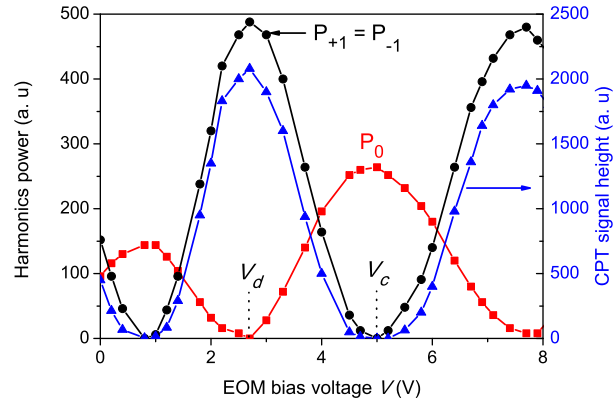}
\caption{Optical power in the carrier (squares), first-order sidebands (circles) at the output of the MZ EOM and height of the CPT signal (triangles) versus the EOM bias voltage $V$. }
\label{fig:tf-eom}
\end{figure}

\begin{figure}[htb]
\centering
\includegraphics[width=0.65\linewidth]{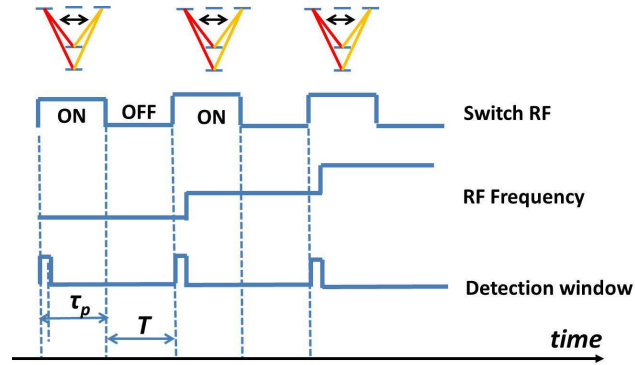}
\caption{Pulsed sequence to scan and detect the CPT Ramsey fringes. The CPT pumping time $\tau_p$ is typically 1-2 ms. The Ramsey time $T$ is of a few ms.}
\label{fig:Ramsey-Sequence}
\end{figure}

Figure \ref{fig:tf-eom} displays the optical harmonics power (height of the peaks at the output of the FPI) and the CPT signal (in continuous regime) at the output of the MZ EOM  versus the bias voltage $V$ when the RF power is continuously activated. The CPT signal height is found to be maximized for the so-called \emph{dark point} $V = V_d =$ 2.6 V where the optical carrier is optimally rejected and the power in first-order optical sidebands is maximized. At the opposite case, the CPT signal vanishes for $V = V_c$ = 5 V where the optical carrier is maximized and first-order optical sidebands power minimized. We checked that when no RF modulation is applied, the output optical power is minimized (carrier suppression point) for $V = V_d$. According to the transfer functions reported above, we implemented the Ramsey sequence described in Fig. \ref{fig:Ramsey-Sequence} to make the atoms interact with a sequence of optical pulse trains. This pulsed interaction scheme is inspired from \cite{Zanon:PRL:2005}.

A first pulse allows to pump the atoms in the CPT state during a time $\tau_p$. For this purpose, the RF power is switched on using a microwave switch (Mini-Circuits ZASW-2-50-DR+) and the EOM bias voltage is adjusted to $V_d$ to reject the optical carrier. Then, the RF power is turned off while the EOM bias voltage is kept to $V_d$ to switch off the light. During this period, the atoms evolve freely in the \emph{dark} for a time $T$. The next pulse is used both for the atomic signal detection and pumping again the atoms in the CPT state. The EOM-based light switching time was measured to be $<$ 10 $\mu$s. The signal is typically detected a time $\tau_d =$ 10 $\mu$s after the pulse trigger and averaged on a measurement time $t_m$ = 25 $\mu$s. The Ramsey time $T$ is ultimately limited by the microwave coherence relaxation time $T_2$ in the CPT cell.
\section{Experimental results}\label{sec:results}
\subsection{Detection of CPT Ramsey fringes}
\begin{figure}[htb]
\centering
\includegraphics[width=0.75\linewidth]{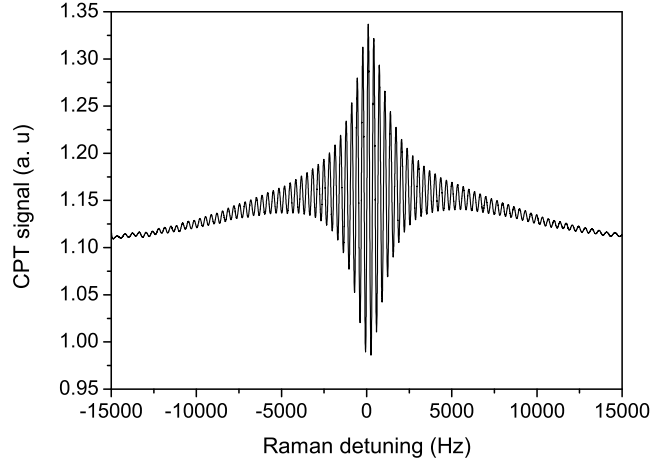}
\caption{166-Hz width Ramsey fringes with a contrast of 33 \%. Experimental conditions: $T$ = 3 ms, $\tau$ = 2 ms, $P_L$ = 3.8 mW.}
\label{fig:ramsey-largescan}
\end{figure}

\begin{figure}[htb]
\centering
\includegraphics[width=0.75\linewidth]{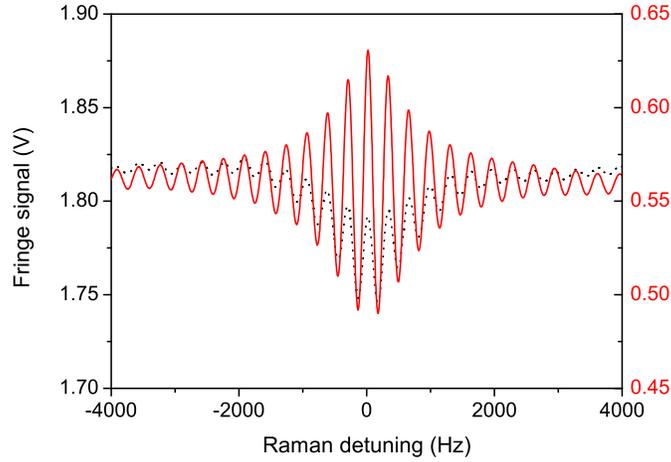}
\caption{Comparison of Ramsey fringes obtained with standard circular polarization (dotted line) and push-pull optical pumping (solid line). Experimental conditions: $T$ = 3 ms, $\tau$ = 2 ms.}
\label{fig:circ-vs-ppop}
\end{figure}

Fig. \ref{fig:ramsey-largescan} reports, on a large span of 30 kHz, Ramsey fringes detected with $\tau_p$ = 2 ms and $T$ = 3 ms for a total laser input power of 3.8 mW. The central fringe exhibits a linewidth $\Delta \nu$ = $1/(2T)=$ 166 Hz with a contrast $C$ of 33 \%. In order to highlight the efficiency of the PPOP technique, we compare on Fig. \ref{fig:circ-vs-ppop} Ramsey fringes obtained with standard circular polarization and push-pull optical pumping for identical experimental conditions. The contrast of the central fringe is found to be 31 \% with push-pull interaction scheme whereas it is only 2.8 \% with circular polarization.

\subsection{Impact on the fringe signal of some experimental parameters}
\begin{figure}[h!]
\centering
\subfigure[]{\includegraphics[width=0.47\linewidth]{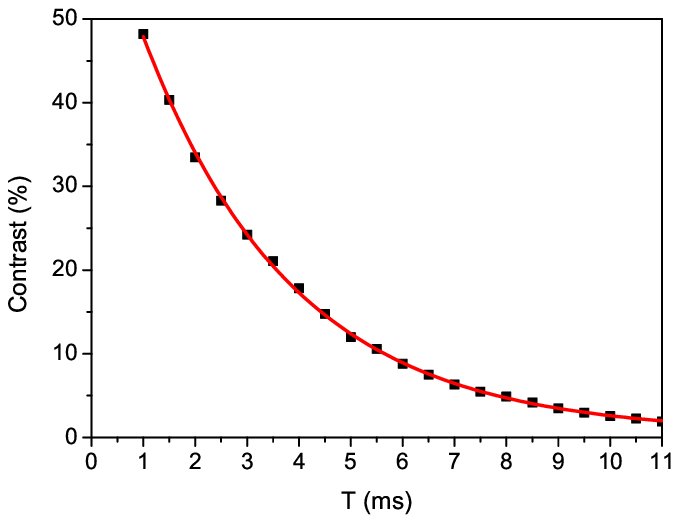}
\label{fig:contrast-vs-T}} \hfill
\subfigure[]{\includegraphics[width=0.47\linewidth]{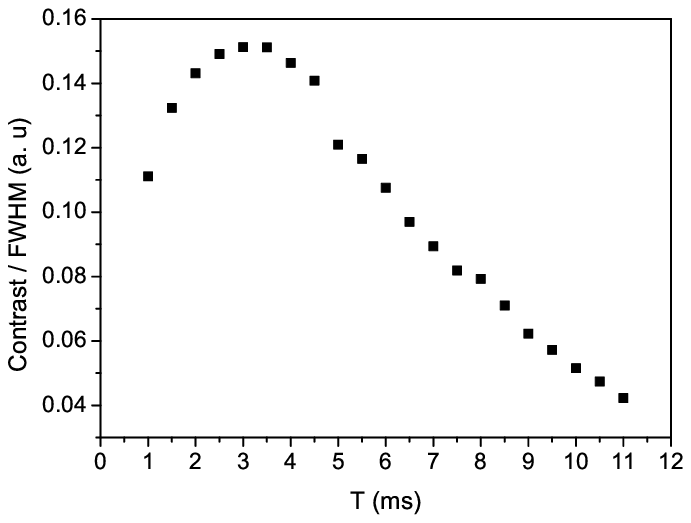}
\label{fig:contrastlarg-vs-T}}  \caption{Contrast (a) and contrast/linewidth ratio (b) of the central Ramsey fringe versus the dark time $T$. The solid line on (a) is an exponential decay function fit to experimental data. Experimental conditions are: $\tau$ = 2 ms, $P_L$ = 3 mW.  \label{fig:5abc}}
\end{figure}

\begin{figure}[htb]
\centering
\includegraphics[width=0.7\linewidth]{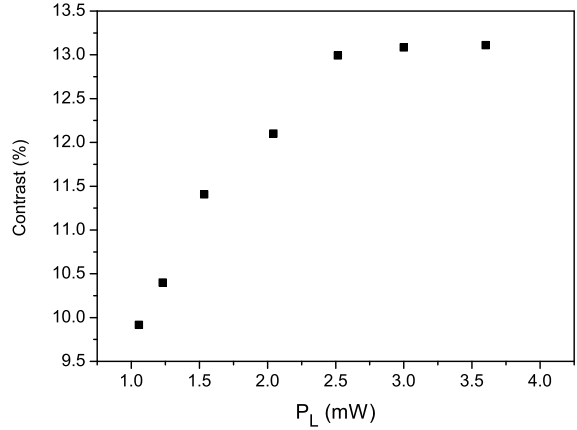}
\caption{Contrast of the central fringe versus the laser power $P_L$. Experimental conditions are: $T$ = 5 ms, $\tau_p$ = 2 ms.}
\label{fig:contrast-vs-pl}
\end{figure}

Figures \ref{fig:contrast-vs-T} and \ref{fig:contrastlarg-vs-T} show the contrast and the contrast/linewidth ratio of the central Ramsey fringe versus the Ramsey time $T$. We note that contrasts as high as 50 \% and 24.2 \% are obtained for $T$ = 1 ms and $T$ = 3 ms respectively. For $T$ = 10 ms (fringe linewidth = 50 Hz), the contrast of the Ramsey fringe is still significant (3 \%). Experimental data on Fig. \ref{fig:contrast-vs-T} are well fitted by an exponential decay function with a time constant of about 3 ms. This value can be interpreted as an experimental estimation of the CPT hyperfine coherence lifetime $T_2$. The ratio contrast/linewidth is maximized for $T = T_2$ = 3 ms. This result indicates that a Ramsey time $T = T_2$ is a good trade-off between a large signal and a narrow fringe for atomic clocks applications, as already observed in \cite{Micalizio:2012} and \cite{Guerandel:IM:2007}.\\
Figure \ref{fig:contrast-vs-pl} plots the contrast of the central fringe versus the laser power $P_L$ for $T$ = 5 ms and $\tau_p$ = 2 ms.
\begin{figure}[htb]
\centering
\includegraphics[width=0.7\linewidth]{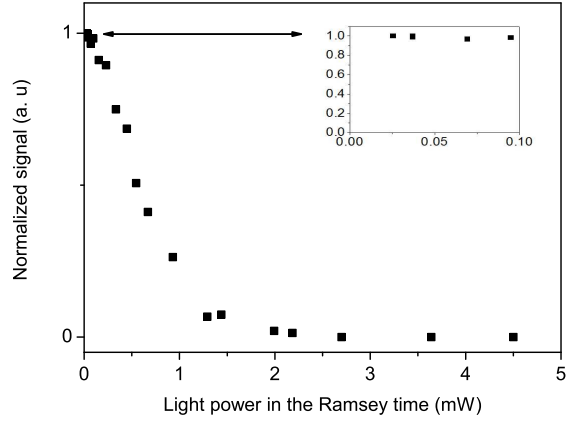}
\caption{CPT fringe peak-peak signal versus laser power seen by the atoms in the Ramsey time $T$. The CPT signal for the lowest power point is normalized to 1. The inset shows a zoom on first points of the graph.}
\label{fig:fringe-noir}
\end{figure}

The contrast of the resonance is increased with increased laser power until $P_L \simeq$ 3 mW before reaching a plateau. For $P_L$=1 mW, the contrast is found to be about 10 \%. For comparison, in \cite{Boudot:IM:2009}, a contrast of 4.6 \% was obtained at $T$ = 5 ms for a laser intensity 2 times lower in a similar cell. Contrary to the CW regime where the CPT linewidth broadens with increased laser power, we measured in the pulsed case that the fringe linewidth does not depend on the laser power and narrows as $1/(2T)$.\\

A possible drawback of the used MZ EOM is to exhibit a relatively low extinction ratio of only 24dB. In order to estimate its effect (see Fig. \ref{fig:fringe-noir}), we degraded intentionally the light extinction by changing the EOM bias voltage $V$ during the free evolution time $T$. The laser power during the pumping time is 3.6 mW. Maximum and minimum optical power seen by the atoms in the dark time are respectively 5.46 mW and 25.6 $\mu$W, limited by the EOM extinction ratio. Note that MZ EOMs with extinction ratio of 60 dB are commercially available. Such devices could allow to reduce greatly the minimum optical power to a level of a few nW during the time $T$. The CPT fringe signal is measured to decrease rapidly when the laser power in the time $T$ is increased. The signal is found to vanish completely for $P_L >$ 1 mW. Interestingly, for a residual light power $<$ 100 $\mu$W, a plateau seems to be reached indicating that a better light extinction does not bring a relevant improvement on the clock signal. Nevertheless, we expect that a total light extinction during the time $T$ is of great importance and to be preferred in order to minimize light shift effects for atomic clocks applications.

\section{Discussion and summary}

The frequency stability of atomic clocks is currently expressed by the Allan deviation $\sigma_y (\tau)$. Assuming the photon shot noise as the only noise source, the shot-noise limit of the clock relative frequency stability can be estimated by:
\begin{equation}
\sigma_y(\tau)\simeq\frac{\Delta \nu}{\nu_0}\frac{1}{C}\sqrt{\frac{h\nu}{P_{o}}}\sqrt{\frac{T_c}{t_m\tau}},
\label{eq}
\end{equation}
where $\Delta \nu$ is the fringe width, $\nu_0$ is the clock frequency, $C$ is the contrast of the resonance, $h\nu$ is the energy of a single photon, $P_0$ is the laser power at the output of the cell, $T_c = T + \tau_p $ is the clock cycle time, $t_m$ is the length of the detection window and $\tau$ is the averaging time of the measurement. From the clock signal shown in Fig. \ref{fig:ramsey-largescan}, Eq. (\ref{eq}) yields $\sigma_y(\tau) \simeq 9 \times 10^{-15}$ at 1 s averaging time. This value is about one order of magnitude better than shot-noise estimations in the CW regime \cite{Liu:PRA:2013}. In real life, this result will be degraded by various noise mechanisms such as the detector noise, the laser intensity and frequency noise or the oscillator phase noise through Dick effect \cite{Dick}. In a first step, we estimate that demonstrating an experimental frequency stability better than 10$^{-12}$ at 1 s integration time with our system is a reasonable objective.\\

In summary, a simple-architecture MZ EOM-based laser system was developed to detect for the first time high-contrast CPT Ramsey fringes by combination of PPOP and Ramsey interrogation.The MZ EOM is used to operate simultaneously as an optical sidebands generator and a fast optical switch. Ramsey fringes with a linewidth in the 50-200 Hz range with contrasts of 3 to 50 \% were detected. The impact of different experimental parameters on the clock fringe was reported. This system could be of great interest for the development of a compact and high-performance CPT-based vapor cell frequency standard.
\section*{Acknowledgments}\label{sec:acknow}
This work was supported by Agence Nationale de la Recherche (ANR) and D\'el\'egation G\'en\'erale de l'Armement in the frame of ISIMAC project (ANR-11-ASTR-0004). X. Liu PhD thesis is funded by R\'egion de Franche-Comt\'e.
\end{document}